\documentclass[twocolumn,showpacs,preprintnumbers,pra]{revtex4}%
\usepackage{amssymb}
\usepackage{graphicx}
\usepackage{dcolumn}
\usepackage{bm}
\usepackage{epsfig}
\usepackage{amsmath}
\usepackage{amsfonts}%
\begin{document}
\title{Hydrodynamic behavior in expanding thermal clouds of $^{87}$Rb}
\author{I.~Shvarchuck,$^{1}$ Ch.~Buggle,$^{1}$ D.S.~Petrov,$^{1,2}$ M.~Kemmann,$^{1}$
W.~von~Klitzing,$^{1}$ G.V.~Shlyapnikov,$^{1,2}$ and J.T.M.~Walraven$^{1}$}
\affiliation{${~}^{1}$ FOM Institute for Atomic and Molecular Physics Kruislaan 407\mbox{,}
1098 SJ Amsterdam\mbox{,} The Netherlands, ${~}^{2}$ Russian Research Center,
Kurchatov Institute, Kurchatov Square, 123182 Moscow, Russia }
\date{\today}

\begin{abstract}
We study hydrodynamic behavior in expanding thermal clouds of $^{87}$Rb
released from an elongated trap. At our highest densities the mean free path
is smaller than the radial size of the cloud. After release the clouds expand
anisotropically. The cloud temperature drops by as much as 30 \%. This is
attributed to isentropic cooling during the early stages of the expansion. We
present an analytical model to describe the expansion and to estimate the
cooling. Important consequences for time-of-flight thermometry are discussed.

\end{abstract}
\maketitle

\section{Introduction}

The anisotropic expansion of a condensate after release from a trap is one of
the best known features of the Bose-Einstein condensed state
\cite{Anderson95,davis95}. The anisotropy arises because the condensate
expands most rapidly in directions where it was originally most confined. The
interest in this phenomenon is further growing, in particular since the
observation of anisotropic expansions in non-condensed Bose gases
\cite{Shvar02,Gerbier03} and in degenerate Fermi gases
\cite{Ohara02,Bourdel03}.

Anisotropic expansions are indicative for hydrodynamic behavior. It is well
known that Thomas-Fermi condensates can be described by the classical Euler
equation for potential flow of a non-viscous gas \cite{Pitaevskii03}.
Therefore, they behave hydrodynamically even at very low densities. For
classical clouds the situation is density dependent. At low densities, where
the mean free path is large compared to the size of the cloud (collisionless
regime), the expansion proceeds under free flow conditions (free expansion).
The motion of the individual atoms is described by a single-particle
Hamiltonian and the expansion is isotropic. Reducing the mean free path to a
value smaller than the dimension of the cloud allows the introduction of a
hydrodynamic field and leads to a crossover to hydrodynamic behavior
(hydrodynamic expansion). Little difference is to be expected between the
expansion of a condensate and that of a fully hydrodynamic thermal cloud
\cite{Kagan97}. Both the collisionless and the hydrodynamic regime were
studied theoretically (see \cite{Kagan97,Griffin97,Guery99,Khawaja2000} and
references therein). Also the influence of mean-field effects \cite{Guery02}
and the crossover between the two regimes were analyzed theoretically
\cite{Pedri03} and numerically \cite{Wu98}.

It is important to understand the crossover to hydrodynamic behavior in
thermal clouds. From the fundamental point of view it is important to quantify
the hydrodynamic properties as these affect the coupling between condensates
and thermal clouds. From the experimental point of view it is vital for the
correct interpretation of time-of-flight absorption images of dense atomic
clouds. Previously the crossover regime in thermal clouds was probed in
experiments at MIT with a dense gas of $^{23}$Na atoms \cite{Stamp98} and at
ENS using cold metastable triplet $^{4}$He \cite{Leduc02}. In Amsterdam the
crossover regime was observed in experiments with $^{87}$Rb \cite{Shvar02}.
Very pronounced hydrodynamic conditions were recently reached by exploiting a
Feschbach resonance in fermionic gases
\cite{Ohara02,Bourdel03,Jochim02,Gupta03,Regal03}. Hydrodynamic behavior as
observed in collective excitations is reviewed in
refs.\cite{Pitaevskii03,Dalfo99}.

In this paper we focus on hydrodynamic behavior as observed in the expansion
of dense thermal clouds of $^{87}$Rb, extending a brief analysis presented
earlier in the context of the BEC formation experiments in Amsterdam
\cite{Shvar02}. The clouds are prepared in an elongated trap at a temperature
$T_{0},$ just above the critical temperature for Bose-Einstein condensation.
At the highest densities the mean free path is less than the radial size of
the cloud. After release from the trap the clouds expand anisotropically and
their temperature drops by as much as $30\%$. The behavior is intermediate
between that expected for collisionless clouds, where cooling is absent, and
pure hydrodynamic behavior, where the gas cools to vanishing temperatures.

We show that the expansion in axial direction is similar to that of a
collisionless cloud at a temperature $T_{z}<T_{0}$. This `axial' temperature
can be identified with the temperature $T_{\ast}$ reached at the moment when
the expansion ceases to be hydrodynamic and the cooling stops. Radially, the
expansion proceeds faster than that expected for a collisionless cloud and can
be characterized by a `radial' temperature $T_{\rho}>T_{0}$. For our
conditions, the mean field of elastic interaction contributes $\sim20\%$ to
the total energy in the trap center. We show that this only has a minor effect
$(3\%)$ on the expansion behavior. The consequences for time-of-flight
thermometry are discussed.

\section{Experiment}

In our experiments we load a magneto-optical trap with approximately 10$^{10}$
atoms from the source described in \cite{Dieckmann96}. After optical pumping
to the $|S_{1/2},F=2,m_{F}=2\rangle$ state typically $4\times10^{9}$ atoms are
captured in a Ioffe-Pritchard quadrupole magnetic trap. Then the gas is
compressed and evaporatively cooled to a temperature just above $T_{C}$. The
radio-frequency (rf) evaporation is forced at a final rate of $\dot{\nu}=-433$
kHz/s down to a value $\nu_{1}=740$ kHz, that is $120$ kHz above the trap
minimum $B_{0}=88.6(1)$~$\mu$T as calibrated using atom laser output coupling
\cite{Bloch99}. As the final ramp down rate is $-{\dot{\nu}/(\nu}_{1}{-\nu
_{0})\approx4}$ s$^{-1},$ i.e. slow compared to both axial and radial trap
frequencies $\omega_{z}=2\pi\times20.8(1)$ s$^{-1}$ and $\omega_{\rho}%
=2\pi\times477(2)$ s$^{-1}$, the evaporation proceeds quasi-statically and
yields a sample characterized by a single uniform temperature and an
equilibrium shape \cite{Note12}. The preparation procedure is completed by
$20$ ms of plain evaporation at rf-frequency $\nu_{1}$. This procedure leaves
us with $N=3.5(3)\times10^{6}$ atoms at density $n_{0}=3.6(6)\times10^{14}$
cm$^{-3}$ in the trap center and temperature $T_{0}$ $=1.17(5)~\mu$K.

\subsection{Knudsen criterion}

To establish the collisional regime we calculate the mean free path and the
atomic collision rate. The mean free path in the trap center is given by the
usual expression for a uniform gas \cite{Chapman70} at density $n_{0}$,
\begin{equation}
\lambda_{0}=\frac{1}{\sqrt{2}n_{0}\sigma}\approx3\mathrm{\,\mu m,}\label{mfp}%
\end{equation}
where $\sigma=8\pi a^{2}$ is the elastic scattering cross-section in the
s-wave limit with $a=98.98(4)a_{0}$ the scattering length \cite{Kempen02}. The
atomic collision rate in the trap center is \cite{Chapman70}%
\begin{equation}
\tau_{c}^{-1}=\sqrt{2}n_{0}\bar{\upsilon}_{th}\sigma\approx6000\,\mathrm{s}%
^{\mathrm{\ -1}},\label{eq3}%
\end{equation}
with $\bar{\upsilon}_{th}=(8k_{B}T_{0}/\pi m)^{1/2}$ as the thermal velocity.

The gas behaves as a hydrodynamic fluid if the mean free path is much smaller
than the relevant sample size (Knudsen criterion). Defining the axial $\left(
l_{z}\right)  $ and the radial $\left(  l_{\rho}\right)  $ size parameters of
the density profile in a harmonic trap, see Eq.(\ref{T0}), the Knudsen
criterion can be expressed as%
\begin{equation}
\frac{\lambda_{0}}{l_{i}}\simeq\omega_{i}\tau_{c}\ll1,
\label{Knudsen criterion}%
\end{equation}
with $i\in\{\rho,z\}.$ For the axial direction the Knudsen criterion is very
well satisfied, $\omega_{z}\tau_{c}\approx0.02$. For the radial direction we
calculate $\omega_{\rho}\tau_{c}\approx0.5$. In this direction we operate in
the middle of the crossover range between the collisionless and hydrodynamic regimes.

\subsection{Time-of-flight analysis}

In the crossover between hydrodynamic and collisionless conditions the
time-of-flight analysis is non-trivial. Unlike in fully collisionless clouds,
the velocity of the individual atoms is not conserved because the gas cools as
it expands. Unlike in fully hydrodynamic clouds, cooling will only proceed
during a finite period. Obviously, if the temperature drops during the
expansion the question arises how to properly extract the temperature of the
cloud from a time-of-flight absorption measurement.

\begin{figure}[ptb]
\epsfxsize=\hsize
\epsfbox{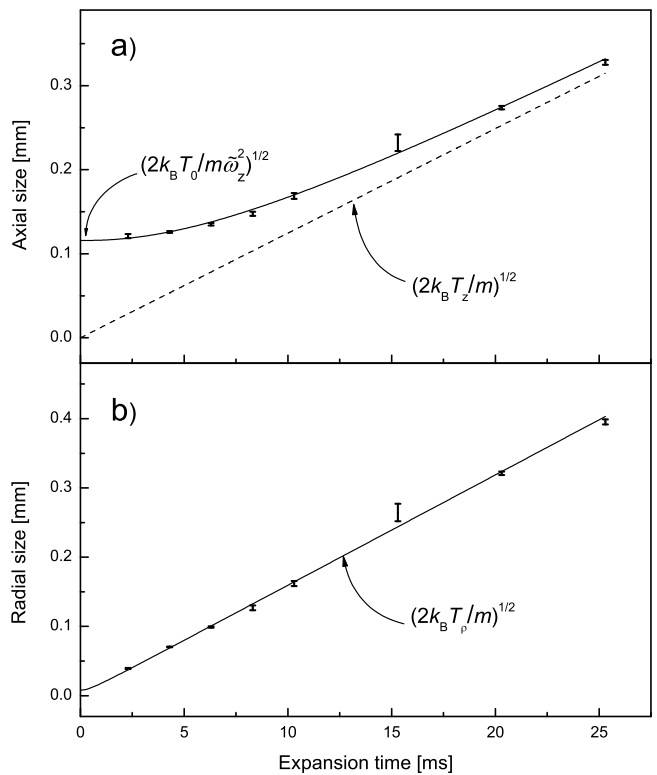}
\caption{Expansion measurements for (a) axial and (b) radial direction. The
error bars represent two standard deviations. The solid lines represent
Eqs.~(\ref{eq8}) and (\ref{eq8a}) with $l_{z}(t_{\ast})=116\,\mu$m,
$T_{z}=0.83$ $\mu$K and $T_{\rho}=1.35$ $\mu$K. Note the difference in
vertical scale for the two panels. The dashed line represents the asymptotic
expansion behavior in axial direction. As the initial radial size is very
small the radial expansion is already asymptotic by the time the first data
point is taken.}%
\label{fig:HDtemperature}%
\end{figure}

In Fig.\thinspace\ref{fig:HDtemperature} we plot the measured axial and radial
cloud sizes, $l_{z}(t)$ and $l_{\rho}(t)$, as a function of expansion time
$t$. All data were collected during a single run within $2.5$ hours, keeping
track of some drift in the ofset field \cite{Note7}. Each data point
corresponds to the average of about $20$ measurements, with the error bars
representing the standard deviation, typically $2\%$ of the average value. The
cloud sizes were determined with the usual procedure (see for instance
ref.~\cite{Ketterle99}), i.e., the expression for the column density of an
ideal Bose gas trapped in a harmonic potential%
\begin{equation}
n_{2}(z,\rho)=n_{20}g_{2}[De^{-\left[  z/l_{z}(t)\right]  ^{2}-\left[
\rho/l_{\rho}(t)\right]  ^{2}}]/g_{2}[D] \label{ColumnDensity}%
\end{equation}
is fitted, after transformation to optical density, to the images
\cite{Note8}. With this procedure we obtain values for the sizes $l_{z}(t)$
and $l_{\rho}(t)$, the degeneracy parameter (fugacity) $D$ and the peak column
density $n_{20}$ \cite{Note11}. We use the notation $g_{a}[x]=\sum
_{l=1}^{\infty}x^{l}/l^{a}.$ The fugacity provides together with the initial
sizes a self-calibrating method for the total atom number provided the average
trap frequency $\bar{\omega}=(\omega_{\rho}^{2}\omega_{z})^{1/3}$ is known,%
\begin{equation}
N=g_{3}[D]\left(  \frac{m\,\bar{\omega}}{2\hbar}\right)  ^{3}l_{z}%
^{2}(0)l_{\rho}^{4}(0). \label{NumberCalibration}%
\end{equation}
In practice only the axial size $l_{z}(0)$ is used because the aspect ratio is
accurately known. The measured peak column density $n_{20}$ is not used in our
analysis \cite{Note9}.

Due to the presence of the elastic interactions between the atoms the density
distribution will be slightly broadened and deformed \cite{Goldman81,Guery02}.
Calculating the variance of the distribution $\left\langle z^{2}\right\rangle
$ using the recursive expression for the density to first order in mean field
$U_{\text{mf}}(\mathbf{r})=2gn(\mathbf{r})$, leads to
\begin{equation}
\frac{1}{2}m\omega_{z}^{2}l_{z}^{2}(0)\simeq kT_{0}+E_{\text{mf}},
\label{T0Emf}%
\end{equation}
where $E_{\text{mf}}=g\int n^{2}(\mathbf{r})d\mathbf{r}/\int n(\mathbf{r}%
)d\mathbf{r}$ is the trap averaged interaction energy with $g=(4\pi\hbar
^{2}/m)a$ the interaction coupling constant \cite{Note5}. The variance was
related to the size parameter using $\left\langle z^{2}\right\rangle =\frac
{1}{2}l_{z}^{2}(0)g_{4}[D]/g_{3}[D]$ \cite{Note6}. Equivalently, treating the
mean field as an effective potential we may write
\begin{equation}
\frac{1}{2}m\tilde{\omega}_{z}^{2}l_{z}^{2}(0)=kT_{0}, \label{T0}%
\end{equation}
where $\tilde{\omega}_{z}$ represents a `dressed' trap frequency that
reproduces, for an ideal gas at temperature $T_{0}$, the same cloud size,
\begin{equation}
\tilde{\omega}_{i}^{2}=\omega_{i}^{2}(1-\xi), \label{DressedFrequencies}%
\end{equation}
where $\xi=E_{\text{mf}}/(kT_{0}+E_{\text{mf}})\approx0.03$ \cite{Note5}.

To describe the expansion behavior analytically we introduce a schematic model
in which the expansion is treated as purely hydrodynamic up to time
$t=t_{\ast}$ and as purely collisionless beyond this point. At $t_{\ast}$ the
density has dropped to the level that no further collisions take place and the
atomic velocities remain frozen. The axial expansion is represented by%
\begin{equation}
l_{z}\left(  t\right)  \simeq\lbrack l_{z}^{2}\left(  t_{\ast}\right)
+(2k_{B}T_{z}/m)\left(  t-t_{\ast}\right)  {^{2}}]^{1/2}. \label{eq8}%
\end{equation}
The presence of $t_{\ast}$ slightly shifts the asymptote of the expansion
curve. The radial expansion is asymptotic for all times relevant in the
experiment,%
\begin{equation}
l_{\rho}\left(  t\right)  \simeq\lbrack2k_{B}T_{\rho}/m]^{1/2}t. \label{eq8a}%
\end{equation}
In this case the shift of the asymptote is negligible. The parameters $T_{z}$
and $T_{\rho}$ represent apparent axial and radial temperatures corresponding
to the asymptotic expansion velocities of the cloud in both directions,
\begin{equation}
s_{i}=\mathop {\lim }\limits_{t\rightarrow\infty}\dot{l}_{i}\left(  t\right)
=(2k_{B}T_{i}/m)^{1/2}, \label{eq9}%
\end{equation}
with $i\in\{\rho,z\}.$ Note that Eqs.(\ref{eq8}) and (\ref{eq8a}) reduce to
the usual expressions for isotropic expansion of fully collisionless thermal
clouds in the absence of a mean field when $t_{\ast}\rightarrow0$ with
$T_{z}=T_{\rho}=T_{0}$ (see e.g. ref.\cite{Ketterle99}).

\section{Results}

Fitting Eq.(\ref{ColumnDensity}) to our data, the degeneracy parameter was
verified to be constant during the expansion to within experimental error,
$D=0.95(4)$. Once this was established we determined the cloud sizes by
refitting all data with a fixed value $D=0.95.$ The results are shown in
Fig.\thinspace\ref{fig:HDtemperature} (solid bars). Fitting Eq.(\ref{eq8}) to
the results for the axial sizes we obtain the initial axial size
$l_{z}(0)\simeq l_{z}(t_{\ast})=116(2)\,\mu$m and the `axial temperature'
$T_{z}=0.83(4)\,\mu$K. The fit is shown as the solid line in Fig.\thinspace
\ref{fig:HDtemperature}a and is insensitive to any reasonable choice of
$t_{\ast}$. Fitting Eq.(\ref{eq8a}) to the radial data we obtain the solid
line in Fig.\thinspace\ref{fig:HDtemperature}b, which corresponds to $T_{\rho
}=1.35(6)\,\mu$K \cite{Note2}. For all these results statistical errors are
negligible. The quoted errors represent the uncertainty in the determination
of the fugacity.

From the initial axial size we calculate with Eq.(\ref{T0}) $T_{0}%
=1.17(5)\,\mu$K. Then, the central density $n_{0}=3.6(6)\times10^{14}$
cm$^{-3}$ follows with
\begin{equation}
n_{0}=g_{3/2}[D]/\Lambda_{0}^{3}, \label{CentralDensity}%
\end{equation}
where $\Lambda_{0}=[2\pi\hbar^{2}/mkT_{0}]^{1/2}$ is the thermal wavelength at
temperature $T_{0}$. Using Eq.(\ref{DressedFrequencies}) to account for the
mean field broadening we calculate with Eq.(\ref{NumberCalibration})
$N=3.5(3)\times10^{6}$ atoms. The error bar reflects the strict conditions on
the atom number imposed by a known fugacity. We return to systematic errors in
the section on thermometry.

\begin{figure}[ptb]
\epsfxsize=\hsize
\epsfbox{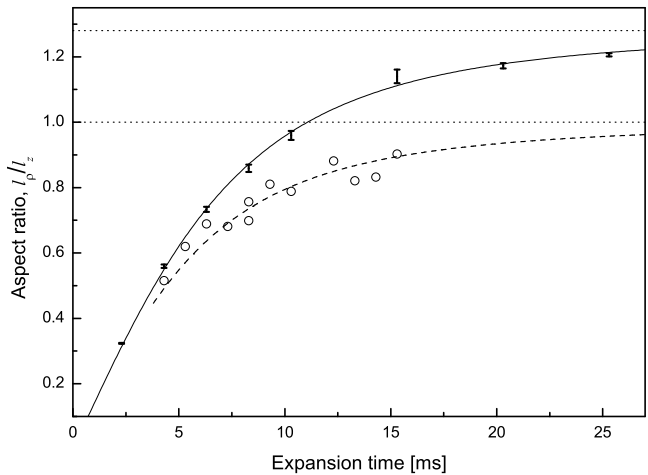}
\caption{Aspect ratio of a hydrodynamically expanding cloud as function of
expansion time. The error bars represent two standard deviations. The change
from a cigar- to a pancake-like shape is evident as the data points cross the
value of $l_{\rho}/l_{z}=1$. The open circles represent low-density clouds
expanding isotropically. The solid and dashed lines represent fits of
Eq.(\ref{eq13}) to the data. }%
\label{fig:AspectRatio}%
\end{figure}

The results presented here indicate a slightly decelerated expansion in axial
direction, $T_{z}/T_{0}=0.71(2),$ and a slightly accelerated expansion in
radial direction, $T_{\rho}/T_{0}=1.15(3)$. This corresponds to an `inversion'
of the aspect ratio, which is demonstrated in Fig.\thinspace
\ref{fig:AspectRatio} by plotting the aspect ratios for the same data set as
used in Fig.\thinspace\ref{fig:HDtemperature}. At $t=12$ ms of expansion the
cloud shape crosses over from a cigar-shape to a pancake-shape. The solid line
represents a fit to the expansion model to be discussed below.

For collisionless samples the expansion is expected to be isotropic. This was
verified by reducing the density by a factor of $30$ (open circles in
Fig.\thinspace\ref{fig:AspectRatio}). In this case the expansion is indeed
isotropic (dashed line), $s_{z}/s_{\rho}=1.02(4)$.

\section{Expansion model}

To interpret our results for $T_{0},$ $T_{\rho}$ and $T_{z}$ we divide the
expansion in two stages. During the first stage $\left(  t<t_{\ast}\right)  $
the expansion is treated as purely hydrodynamic and is described by scaling
theory \cite{Kagan97,Pedri03}. All data are taken during the second stage
$\left(  t>t_{\ast}\right)  $ for which the expansion is treated as collisionless.

\subsection{Hydrodynamic stage}

During the hydrodynamic stage $\left(  t<t_{\ast}\right)  $ we treat the
expansion as isentropic, i.e., the gas cools while converting random motion
into directed motion just as in the supersonic expansion of an atomic beam
\cite{Scoles92}. As for isentropic expansions the degeneracy parameter $D$ is
conserved \cite{Entropy} we find, using Eq.(\ref{CentralDensity}), that the
temperature decreases according to
\begin{equation}
T(t)=T_{0}[n(t)/n_{0}]^{2/3}.
\end{equation}

Turning to scaled size parameters, $b_{i}\left(  t\right)  \equiv
l_{i}(t)/l_{i}(0)$ with $i\in\{\rho,z\},$ the density ratio is conveniently
written as
\begin{equation}
\frac{n(t)}{n_{0}}=\frac{1}{b_{\rho}^{2}(t)b_{z}(t)}. \label{density ratio}%
\end{equation}
We note that for our elongated clouds $(\omega_{z}/\omega_{\rho}\ll1)$ the
axial size remains practically unchanged during the early stages of the
expansion. Therefore, setting $b_{z}=1$ in Eq.(\ref{density ratio}), the
initial $(t\ll1/\omega_{z})$ isentropic drop in temperature can be written as%
\begin{equation}
T(t)/T_{0}=1/b_{\rho}^{4/3}(t). \label{t-short}%
\end{equation}
Here $b_{\rho}(t)$ satisfies the scaling equations for expanding hydrodynamic
thermal clouds \cite{Kagan97} in the presence of a mean field \cite{Pedri03},%
\begin{subequations}
\begin{align}
\ddot{b}_{\rho}  &  =(1-\xi)\frac{\omega_{\rho}^{2}}{b_{\rho}^{7/3}b_{z}%
^{2/3}}+\xi\frac{\omega_{\rho}^{2}}{b_{\rho}^{3}b_{z}},\label{eq10a}\\
\ddot{b}_{z}  &  =(1-\xi)\frac{\omega_{z}^{2}}{b_{z}^{5/3}b_{\rho}^{4/3}}%
+\xi\frac{\omega_{z}^{2}}{b_{\rho}^{2}b_{z}^{2}}. \label{eq10b}%
\end{align}
The Eqs.(\ref{eq10a}) and (\ref{eq10b}) decouple for $t\ll1/\omega_{z}$ since
$b_{z}\simeq1$. In this limit the radial scaling equation can be written as
\end{subequations}
\begin{equation}
\left(  \frac{\dot{b}_{\rho}(t)}{\omega_{\rho}}\right)  ^{2}=\frac{3}{2}%
(1-\xi)[1-1/b_{\rho}^{4/3}(t)]+\xi\lbrack1-1/b_{\rho}^{2}(t)].
\label{b-rho-dot}%
\end{equation}
We then substitute Eq.(\ref{b-rho-dot}) into (\ref{t-short}) and obtain to
first order in ($\dot{b}_{\rho}/\omega_{\rho})^{2}$ the temperature $T_{\ast}$
reached at $t=t_{\ast}$:
\begin{equation}
\frac{T_{\ast}}{T_{0}}\simeq1-\frac{2}{3}\left(  \frac{\dot{b}_{\rho}}%
{\omega_{\rho}}\right)  _{t=t_{\ast}}^{2}. \label{T-inf}%
\end{equation}

We point out that in the limit of very elongated clouds Eq.(\ref{b-rho-dot})
also represents the correct description for a fully hydrodynamic expansion.
Then, we may write for the asymptotic expansion velocity in radial direction
\begin{equation}
\mathop {\lim }\limits_{t\rightarrow\infty}\frac{\dot{b}_{\rho}(t)}%
{\omega_{\rho}}=\frac{1}{\omega_{\rho}}\frac{s_{\rho}}{l_{\rho}\left(
0\right)  }=\sqrt{(1-\xi)\frac{T_{\rho}}{T_{0}}}.
\end{equation}
Hence, comparing with the asymptotic value of Eq.(\ref{b-rho-dot}) we conclude
that the following inequality should hold:
\begin{equation}
1\leq T_{\rho}/T_{0}\leq3/2+\xi. \label{t-rho-inequality}%
\end{equation}

Returning to our experimental conditions we emphasize that the duration of the
hydrodynamic stage will be very brief because the instantaneous mean free path
grows \textit{quadratically} with $b_{\rho}$ in these elongated clouds,
\begin{equation}
\lambda(t)/\lambda_{0}=b_{\rho}^{2}(t), \label{mfp-ratio}%
\end{equation}
as follows with Eqs.(\ref{mfp}) and (\ref{density ratio}) \cite{Note4}.
Roughly speaking $t_{\ast}$ is reached when the mean free path equals the
radial size of the cloud. Therefore, a rough estimate for $t_{\ast}$ can be
obtained by substituting $\lambda(t)=l_{\rho}(t)$ into Eq.(\ref{mfp-ratio})
for $t=t_{\ast}.$ With Eq.(\ref{Knudsen criterion}) this leads to%
\begin{equation}
b_{\rho}(t_{\ast})\simeq1/\tilde{\omega}_{\rho}\tau_{c}. \label{b-rho-t-star}%
\end{equation}
As for $t\lesssim1/\tilde{\omega}_{\rho}$ the radial size of a hydrodynamic
cloud hardly differs from that of a collisionless cloud,
\begin{equation}
b_{\rho}(t)\simeq\left(  1+\tilde{\omega}_{\rho}^{2}t^{2}\right)  ^{1/2},
\label{b-rho-collisionless}%
\end{equation}
we find with Eq.(\ref{b-rho-t-star})
\begin{equation}
t_{\ast}\simeq(1/\tilde{\omega}_{\rho})[(1/\tilde{\omega}_{\rho}\tau_{c}%
)^{2}-1]^{1/2}\approx0.6\text{ ms}. \label{t-star-1}%
\end{equation}
A self consistent estimate for our expansion model can be obtained by
combining Eqs.(\ref{b-rho-collisionless}) and (\ref{t-short}) for $t=t_{\ast
},$
\begin{equation}
t_{\ast}\simeq(1/\tilde{\omega}_{\rho})[\left(  T_{0}/T_{\ast}\right)
^{3/2}-1]^{1/2}. \label{t-star-2}%
\end{equation}
However, for this estimate the ratio $T_{\ast}/T_{0}$ should first be
established experimentally.

\subsection{Collisionless stage}

Once the expansion is ballistic ($t>t_{\ast}$) the variance of the axial
$(i=z)$ and radial $(i=\rho)$ velocity components of the expanding gas can be
written as
\begin{equation}
\left\langle v_{i}^{2}\right\rangle =\left\langle u_{i}^{2}\right\rangle
+\left\langle w_{i}^{2}\right\rangle , \label{variance}%
\end{equation}
where $u_{i}$ represents the thermal velocity components of the atoms and
$w_{i}$ the dynamic velocity components of the density distribution due to the expansion.

At the start of the ballistic stage $(t=t_{\ast})$ the thermal velocity
components can be associated with $T_{\ast}$,
\begin{equation}
m\left\langle u_{i}^{2}\right\rangle =k_{B}T_{\ast}\text{ }. \label{mw1}%
\end{equation}
The dynamical velocities due to the overall expansion can be expressed as%
\begin{equation}
m\left\langle w_{i}^{2}\right\rangle =m\left\langle \dot{r}_{i}^{2}%
\right\rangle =(\dot{b}_{i}/\tilde{\omega}_{i})^{2}k_{B}T_{0}=\frac{1}{2}%
m\dot{l}_{i}^{2}. \label{mw2}%
\end{equation}
Here we used the scaling property $\dot{r}_{i}=(\dot{b}_{i}/b_{i})r_{i}$, with
the $r_{i}$ representing the position coordinates in the expanding cloud.
Since for collisionless clouds the $\left\langle v_{i}^{2}\right\rangle $ are
conserved by the time the mean field has vanished, we may write
\begin{equation}
m\left\langle v_{i}^{2}\right\rangle =k_{B}T_{i},
\end{equation}
where the $T_{i}$ are effective axial and radial temperatures that may be
associated with the asymptotic axial and radial expansion velocities $s_{i}$
defined in Eq.(\ref{eq9}).

Substituting Eqs.(\ref{mw1}) and (\ref{mw2}) into Eq.(\ref{variance}) we
obtain for $\xi\ll1$%
\begin{equation}
\frac{T_{\rho}}{T_{0}}=\frac{T_{\ast}}{T_{0}}+\left(  \frac{\dot{b}_{\rho}%
}{\tilde{\omega}_{\rho}}\right)  _{t=t_{\ast}}^{2}+\frac{\xi}{b_{\rho}%
^{2}(t_{\ast})}, \label{T-sub-rho}%
\end{equation}
where the second term on the r.h.s. represents both the hydrodynamic and mean
field contributions to the dynamic motion at $t=t_{\ast}\ll1/\omega_{z}$ and
the third term the mean field contribution to the dynamic motion for
$t>t_{\ast}$ \cite{Note3}. With Eq.(\ref{b-rho-dot}) this results in the
following relation between $T_{0},$ $T_{\rho}$ and $T_{\ast}$ in expanding
elongated thermal clouds%
\begin{equation}
\frac{3}{2}T_{0}+\xi T_{0}=\frac{1}{2}T_{\ast}+T_{\rho}.
\label{energy conservation}%
\end{equation}
This equation is valid for small mean fields provided $t_{\ast}\ll1/\omega
_{z}$ and expresses the energy conservation during the expansion. It implies
\begin{equation}
T_{z}=T_{\ast}. \label{t-z-equals-t-star}%
\end{equation}
This also follows directly by writing in analogy to Eq.(\ref{T-sub-rho})%
\begin{equation}
\frac{T_{z}}{T_{0}}=\frac{T_{\ast}}{T_{0}}+\left(  \frac{\dot{b}_{z}}%
{\tilde{\omega}_{z}}\right)  _{t=t_{\ast}}^{2}, \label{T-sub-Z}%
\end{equation}
taking into account that $(\dot{b}_{z}/\tilde{\omega}_{z})^{2}$ is negligibly
small \cite{Note1}.

\section{Thermometry}

The result (\ref{t-z-equals-t-star}) shows that with our measurement of
$T_{z}$ we directly probe the temperature of elongated clouds at the end of
the hydrodynamic stage. Knowledge of $T_{\ast}$ allows to obtain with
Eq.(\ref{t-star-2}) a self-consistent result for $t_{\ast}$ within our
expansion model. Using $T_{\ast}/T_{0}=0.71(2)$ we calculate $t_{\ast}=0.28$
ms, somewhat smaller than the rough estimate (\ref{t-star-1})

Rewriting (\ref{energy conservation}) we find an increase in the effective
radial temperature
\begin{equation}
\frac{T_{\rho}}{T_{0}}=\frac{3}{2}(1-\frac{1}{3}\frac{T_{\ast}}{T_{0}}%
)+\xi=1.18(2). \label{nicely}%
\end{equation}
Hence $15\%$ of the increase in $T_{\rho}$ is due to the mean field. Note that
Eq.(\ref{nicely}) satisfies inequality (\ref{t-rho-inequality}). Notice
further that the value $T_{\rho}=1.37(6)$~$\mu$K obtained with
Eq.(\ref{nicely}) comes close to the value $T_{\rho}=1.35(6)$~$\mu$K following
directly from the radial expansion.

We found the fitting procedure for determining $T_{0},$ $T_{\rho}$ and
$T_{\ast}$ to be very sensitive for the detailed shape of the fit function.
Choosing a simple gaussian reduces the estimated values for these temperatures
by as much as $25\%$. However, this enormous systematic error does not affect
the corresponding aspect ratios by more than a few parts in a thousand. We
found more indicators that the aspect ratios are more accurately determined
than the absolute values. Interestingly, we find for the aspect ratios
standard deviations of typically $1\%$, i.e. twice as small as for the
absolute size \cite{Note13}. This points to some form of error cancellation.
Also the fit to the aspect ratio is somewhat better than those of the separate plots.

Let us now turn to the results for the aspect ratios as presented in
Fig.\thinspace\ref{fig:AspectRatio}. Using Eqs.(\ref{eq8}), (\ref{eq8a}) and
(\ref{energy conservation}) the evolution of the aspect ratio can be expressed
as
\begin{equation}
\frac{l_{\rho}(t)}{l_{z}(t)}\simeq\frac{\lbrack(3/2+\xi)-1/2\,(T_{\ast}%
/T_{0})]^{1/2}\omega_{z}t}{[1+\xi+(T_{\ast}/T_{0})\omega_{z}^{2}(t-t_{\ast
})^{2}]^{1/2}}, \label{eq13}%
\end{equation}
where we presume $t\gg1/\omega_{\rho}$ as in Eq.(\ref{eq8a}). By construction
this form satisfies energy conservation. In this way our fitting function
stays as close as possible to a fit to a solution of the scaling equations.
Fitting Eq.(\ref{eq13}) to the data using $\xi=0.03$ and $t_{\ast}=0.3$~ms we
obtain $T_{\ast}/T_{0}=0.72(1)$. The fit is shown as the solid line in
Fig.\thinspace\ref{fig:AspectRatio}. The result agrees within experimental
error with that obtained from the axial expansion data but the accuracy is
slightly better. The method lacks the accuracy to extract $\xi$ \cite{Note10}.
The dashed line in Fig.\thinspace\ref{fig:AspectRatio} corresponds to the
collisionless limit of Eq.(\ref{eq13}): $\xi=0$, $t_{\ast}=0$ and $T_{\ast
}=T_{0}$.

Once Eq.(\ref{eq13}) is accepted, time-of-flight information for a single
expansion time suffices for thermometry. The procedure goes in two steps.
First we set $\xi$ and $t_{\ast}$ equal to zero and use Eq.(\ref{eq13}) to
obtain a first estimate for $T_{\ast}/T_{0}.$ With Eq.(\ref{nicely}) $T_{\rho
}/T_{0}$ follows. After $T_{\rho}$ is determined with Eq.(\ref{eq8a}), we have
an estimate for the absolute value $T_{0}$. Together with $n_{0},$ deduced
with Eq.(\ref{CentralDensity}), this allows us to calculate $\xi$ and
$t_{\ast}$. Iterating the procedure once yields all values within the limits
of accuracy of the analysis. Choosing the expansion time sufficiently long
$(t\gg1/\omega_{z})$ the results are very insensitive for the value of
$t_{\ast}$.

Our estimates for the absolute values of $T_{0},$ $T_{\rho}$ and $T_{z}$ are
sensitive for the detailed shape of the clouds. Therefore, deviations from the
Bose shape will result in systematic errors, in particular if the cloud shape
changes during the expansion. Shape deviations can arise from the presence of
the mean field. Also, inhomogeneous isentropic cooling as a result of the
inhomogeneous density profile of our samples can give rise to deviations of
the Bose shape. Further, it may be that our transformation from optical
density to column density gives rise to slight distortions of the cloud shape
as a result of optical pumping or saturation of the detection transition.

In our analysis we did not correct for deviations of the cloud shape from the
Bose distribution. First of all because under our conditions the mean field is
weak and our fits of Eq.(\ref{ColumnDensity}) to the measured column densities
look convincingly. Secondly, because shape deviations produce similar relative
errors in all three temperatures. Therefore, they do not affect the
conclusions and consistency of our analysis as long as the scaling approach
remains valid.

\section{Conclusions}

We studied the behavior of dense elongated clouds of $^{87}$Rb in the
crossover from the collisionless to the hydrodynamic regime. At our highest
densities the mean free path is slightly smaller than the radial size of the
cloud and the expansion is anisotropic. The expansion can be described by a
two stage model in which the expansion is treated as purely hydrodynamic up to
time $t=t_{\ast}$ and as purely collisionless beyond this point. We find that
at the end of the hydrodynamic stage the temperature has dropped substantially
due to isentropic cooling, $T_{\ast}/T_{0}=0.72(1)$. This reflects itself in
an axial expansion that is substantially slower than expected for the
collisionless case, $T_{z}=T_{\ast}$. In accordance with energy conservation
the radial expansion is faster, $T_{\rho}>T_{0}$. The isentropic cooling is
best determined from the aspect ratio. Although the mean field in the trap
center is substantial, $U_{\text{mf}}(0)/kT_{0}=0.23$, it hardly affects the
expansion behavior. Including the mean field in the analysis only affects the
value obtained for $T_{0}$, with $T_{z}$ and $T_{\rho}$ by definition being
unaffected. In our case the mean-field corrections are too small to be
extracted with a fitting procedure but can be calculated accurately. It leads
to systematic errors in $T_{0}$ of order $3\%$ if not included.

Presently it is possible to study the case of strong mean fields by tuning to
a Feschbach resonance
\cite{Inouye98,Cornish98,Ohara02,Bourdel03,Jochim02,Gupta03,Regal03}. It would
be interesting to study the case where anisotropic expansions are to be
expected, but the behavior of the system is dominated by the mean field rather
than the collisional hydrodynamics.\vspace*{0.2cm}

\section{Acknowledgements}

The authors wish to thank Paolo Pedri for stimulating discussions. This work
is part of the research program on Cold Atoms of the Stichting voor
Fundamenteel Onderzoek der Materie (FOM), which is financially supported by
the Nederlandse Organisatie voor Wetenschappelijk Onderzoek (NWO). Further,
the research received support from NWO under project 047.009.010, from INTAS
under project 2001.2344, and from the Russian Foundation for Basic Research (RFBR).


\begin{thebibliography}{99}
\bibitem {Entropy}For a harmonic trap the entropy is given by the expression
$S=Nk_{B}(4g_{4}[D]/g_{3}[D]-\ln D).$
\end{thebibliography}
\end{document}